\documentclass[12pt,preprint]{aastex}
\usepackage{emulateapj5,apjfonts}
\input psfig.tex

\newcommand{\kms}{$\rm {km}~\rm s^{-1}$}
\newcommand{\Msun}{M_\odot}

\begin{document}

\title{VLA Limits for Intermediate Mass Black Holes in Three Globular Clusters}
\author{F.N. Bash\altaffilmark{1}, K. Gebhardt\altaffilmark{1},
W.M. Goss\altaffilmark{2},
P.A. Vanden Bout\altaffilmark{3}}

\altaffiltext{1}{Department of Astronomy, University of Texas at
Austin 1 University Station C1400, Austin, TX 78712;
fnb@astro.as.utexas.edu, gebhardt@astro.as.utexas.edu}

\altaffiltext{2}{National Radio Astronomy Observatory, P.O. Box 0,
Socorro NM 878701; mgoss@nrao.edu}

\altaffiltext{3}{National Radio Astronomy Observatory, 520 Edgemont
Road, Charlottesville, VA 22903; pvandenb@nrao.edu}

\begin{abstract}

The observational evidence for central black holes in globular
clusters has been argued extensively, and their existence has
important consequences for both the formation and evolution of the
cluster. Most of the evidence comes from dynamical arguments, but the
interpretation is difficult, given the short relaxation times and old
ages of the clusters. One of the most robust signatures for the
existence of a black hole is radio and/or X-ray emission. We observed
three globular clusters, NGC6093 (M80), NGC6266 (M62), and NGC7078
(M15), with the VLA in the A and C configuration with a 3-$\sigma$
noise of 36, 36 and 25 $\mu$Jy, respectively. We find no
statistically-significant evidence for radio emission from the central
region for any of the three clusters. NGC6266 shows a 2-$\sigma$
detection. It is difficult to infer a mass from these upper limits due
to uncertainty about the central gas density, accretion rate, and
accretion model.

\end{abstract}


\section{Introduction}

Although we do not understand how the nuclei of galaxies form or why
they have black holes (BH) at their centers, the correlation between
BH mass and bulge velocity dispersion does shed light on their
formation and evolutionary histories (Gebhardt et al. 2000a, 2000b:
Ferrarese and Merritt 2000). A number of different theories (e.g.,
Silk \& Rees 1998; Haehnelt \& Kauffmann 2000; Robertson et al. 2006)
predict a BH mass bulge-velocity-dispersion relation, although they
predict different slopes and intercepts for this relation.
Exploration of the extreme ends of this relationship will help
illuminate the underlying physical model, and in this paper we focus
on the low mass end. Black holes at the low end of the relations, with
masses between 100 and $10^6~\Msun$, are generally referred to as
intermediate-mass black holes (IMBHs). There is significant evidence
that black hole masses from $10^5-10^6~\Msun$ exist from the work of
Barth, Greene \& Ho (2005) and Greene \& Ho (2006). To go to yet
smaller black hole masses, an extrapolation of the correlation between
black hole mass and stellar velocity dispersion suggests studying
stellar systems with velocity dispersions of 10--20~\kms. These
dispersions are characteristic of globular clusters. Whether the
existence of black holes in globular clusters could shed light on the
formation and correlations of supermassive black holes is unknown, but
clearly it is a possibility. Furthermore, the existence of massive
black holes in clusters will have a significant effect on the cluster
evolution. Thus, quantifying the demographics of black holes in
clusters may be related to how supermassive black holes grow, and will
definitely yield useful information about the evolution of
clusters.

Theoretical work suggests that we might expect IMBHs at the centers of
steller systems (Ebisuzaki et al. 2001; Portegies Zwart \& McMillian
2002; Miller \& Hamilton 2002), although it appears to be difficult to
make black holes more massive than 100~$\Msun$.  Gurkan et al. (2004)
suggest that IMBHs may be easy to form through runaway collisions with
massive stars.  Discoveries of BHs in globular clusters have been
claimed --- G1 in M31 (Gebhardt, Rich \& Ho 2002) and M15 (van der
Marel et al. 2002; Gerssen et al. 2002). In fact, the M15 claim has
been made for the past 30 years, starting with the result of Newell,
da Costa \& Norris (1976) and subsequently challenged by Illingworth
\& King (1977). The basic issue is being able to distinguish a rise in
the central mass-to-light ratio being due to either a black hole or
the expected stellar remnants (neutron stars, massive white dwarfs and
solar mass black holes). The most recent M15 result has been
challenged by Baumgardt et al. (2003a). The result in G1 has also been
challenged by Baumgardt et al. (2003b) but Gebhardt, Rich \& Ho (2005)
include additional data and analysis that support the black hole
interpretation. 

There has been two further observations which strongly support the
existence of a black hole in G1. Trudolyubov \& Priedhorsky (2004)
measure X-rays from G1 using the Chandra Observatory, centered to
within 2\arcsec\ of the center of G1. Subsequently, Pooley \&
Rappaport (2006) suggest the X-ray emmission is from accretion onto a
black hole, and Maccarone \& Koerding (2006) point out that if a black
hole is present then a 30 $\mu$Jy radio source may be expected. The
most significant observation comes from Ulvestad, Greene \& Ho (2007)
who find a 28 $\mu$Jy (4.5$\sigma$) emission centered on G1. Other
interpretations are a pulsar wind or a planetary nebula. The pulsar
wind seems unlikely given the age of G1 and the point-like radio
source (an old pulsar would have a large size). A planetary nebula
would show optical emission lines which are not seen in the HST or
Keck spectra of Gebhardt et al. (2003).

Other studies of the existence of black holes in globular clusters have
been less compelling. Colpi, Mapelli, \& Possenti (2003) use indirect
dynamical arguments to suggest a few hundred solar mass black hole in
NGC~6752. McLaughlin et al. (2006) provide an estimate of black hole
in 47Tuc of $900\pm900~\Msun$. To date, there are no published
upper limits of black hole masses that are significantly below that
expected from an extrapolation of the correlation between black hole mass
and stellar velocity dispersion.

While the dynamical arguements strongly support the black hole
interpretation in at least G1, the radio emission provides a clear and
obvious result. Unfortunately, it is difficult to predict the radio
emission from a given black hole mass. The next step is to explore
other globular clusters with a similar setup and deep exposures.

\section{Target Sample and Radio Flux Density Predictions}

We selected three globular clusters for observation.  First, using the
stellar velocity dispersion at the center of M15 (NGC7078) suggests a
revision of the mass of the BH at the cluster's center to 1000
M$_{\sun}$, higher than Maccarone's assumed value of 440 M$_{\sun}$,
making M15 a promising candidate.  Second, noting that Baumgardt et
al. (2004) argue that highly centrally condensed globular clusters, as
seen from their luminosity profiles, are unlikely to harbor central
IMBHs, we selected two globular clusters with large cores that are
more likely to have central BHs.  These clusters, NGC6093 and NGC6266,
also have large central stellar velocity dispersions.

In order to predict radio flux density, the first step is to use an
expected black hole mass. The black hole masses can be estimated using
an extrapolation of the correlations seen in galaxies, namely either
the black hole mass/velocity dispersion or the black hole mass/galaxy
bulge luminosity relations.

A precise prediction of the expected radio flux densities based the
black hole mass is quite uncertain.  Merloni et al. (2003) use radio
flux densities, X-ray luminosities, and measured black hole masses
from both galactic and galaxian black holes to derive a fundamental
plane for the three parameters. They argue that using any one
parameter to predict another is quite uncertain. Unfortunately, X-ray
luminosities do not exist for the three clusters studied
here. Furthermore, the Merloni et al. study do not include any black
holes with masses from 10 to $10^6$, making any use of the fundamental
plane suspect for the three globular clusters. Therefore, instead of
directly using expected black hole mass and measured X-ray luminosity
to predict the radio flux density, we simply use use the location
between the $10^6\Msun$ black holes and the galactic black holes in
the fundamental plane. In this region, the expected 5~Ghz radio power
ranges from $10^{30}-10^{32}$ ergs/s. Indeed, for the G1 radio
emission from Ulvestad, Greene \& Ho (2007) corresponds to $10^{32}$
ergs/s, which is consistent with the measured black hole mass of
$2\times10^4$ from Gebhardt, Rich, \& Ho (2005). Thus, in order to
predict expected flux densities, we adopt this range in radio power
and use the known distances of the globular clusters. A significant
assumption in these estimates is that the physical conditions are
similar; if, for example, the gas density where much lower in globular
clusters, the predicted radio power would be much less.

Alternatively, Maccarone (2004) estimate expected radio emission based
on the expected gas density and the correlation of Merloni et
al. (2003). The gas density in the cluster come from the estimate of
Freire et al. (2001) who use differences in column densities measured
from pulsars in the front and back sides in the globular cluster
47~Tuc. While there is no reason to expect similar gas densities from
cluster to cluster, it is the best measure we have of gas density in a
cluster and therefore we adopt that value. Maccarone (2004) further
assumes that the BH is accreting intra-cluster gas at 0.1 and 1\% of
the Bondi accretion rate. He assumes the BH mass to be 0.1\% of the
globular cluster mass, which he estimates from the cluster's total
luminosity and an assumed mass-to-light ratio, and computes the
expected 5 GHz flux density from the vicinity of the central BH for 15
globular clusters. Six of the globular clusters in Maccaroni's list
lie north of the southern declination limit of the Very Large
Array\footnote{The Very Large Array is a facility of the National
Radio Astronomy Observatory, operated by Associated Universities,
Inc., under a cooperative agreement with the National Science
Foundation.} (VLA) and have an estimated 5 GHz flux density of 40 $\mu
\rm Jy$ or greater (at 1\% of the Bondi rate).  We searched the VLA
archive for observations of the centers of these clusters with noise
levels low enough to have allowed a detection at Maccarone's predicted
levels.  No VLA archive data were found which had the required
sensitivity.

There have been two similar studies to the one presented
here. Maccarone et al. (2005) provide upper limits for omega Cen using
ATCA observations and for M15 from archival VLA observations. De
Rijcke et al. (2006) provide upper limits for 47~Tuc and NGC~6397
based on ATCA observations.

Table 1 shows the computed flux densities for NGC6266, NGC7078, and
NGC6093 at a frequency of 8.6 GHz.  We assume a spectral index,
$\alpha\ = -0.7$ to provide predicted fluxes at 8.6 GHz, the frequency
of our observations.  Our BH masses come from the BH/sigma correlation
(shown in the second column), and we also report those from Maccarone
based on luminosity (the first column).

\section{Observations}

Source positions, integration times on source, beam dimensions and
position angles, and our 3$\sigma$ limits are given in Table 2.

We used the position of the center of M15 determined by Noyola \&
Gebhardt (2006) using the optical surface brightness profile from the
Hubble Space Telescope (HST), which are good to less than 1\arcsec.
We observed that position using the VLA for 7.5 hours on October 13,
2004 in the A configuration at 8.6 GHz ($\lambda$ 3.5 cm), where the
VLA has its maximum sensitivity.  The resulting map has an rms noise
level of 8.5 $\mu \rm Jy$/beam and covers $\approx$ 1 arcmin centered
on the center of M15.  We clearly see the source AC211 reported by
Johnston, Kulkarni, and Goss (1991) about 1.5 arcsec northwest of the
cluster center at with a peak flux density on our map of 144 $\mu \rm
Jy$. We do not detect the other known low mass X-ray binary, M15 X-2
(White \& Angelini 2001), even at 1-sigma. The image also contains the
planetary nebula K 648, for which we get: R.A.  21:29:59.39,
Dec. 12:10:26.46 (J2000).  The measured flux density is 4.2 $\pm$ 0.2
mJy and the deconvolved size is 1.5 x 0.7 arcsec.  We do not see the
pulsar PSR 2127+12A or any other point source near the center of M15
at or above a level of 25 $\mu \rm Jy$ /beam.

The positions of the centers of NGC6093 and NGC6266 also come from
Noyola \& Gebhardt (2006) (with a similar accuracy of around
1\arcsec).  These clusters were observed on August 11, 20, and 26,
2005 using the VLA in the C configuration at 8.6 GHz.  Each map has an
rms noise of 12 $\mu \rm Jy$/beam.  These maps cover an area of 5.4
arcmin centered on each cluster's center.  No radio source is seen at
or above a level of 36 $\mu \rm Jy$/beam near either cluster's center.
In the case of NGC6093 a source is detected 82 arcsec to the SE of the
cluster center at R.A. 16:17:05.00, Dec. -22:59:47.3 (J2000) with a
flux density of 0.32 $\pm$ 0.05 mJy.  The distance from the cluster
center makes it unlikely that this source is related to NGC6266.  No
source is seen convincingly in the image of NGC6266.

Figure 1--3 plots the HST optical images (as used in Noyola \&
Gebhardt) with the VLA radio contours. None of the three clusters show
a significant detection at 8Ghz at the cluster center. However,
NGC~6266 (M62) shows about a 2-$\sigma$ peak at the center.

\vskip 10pt \psfig{file=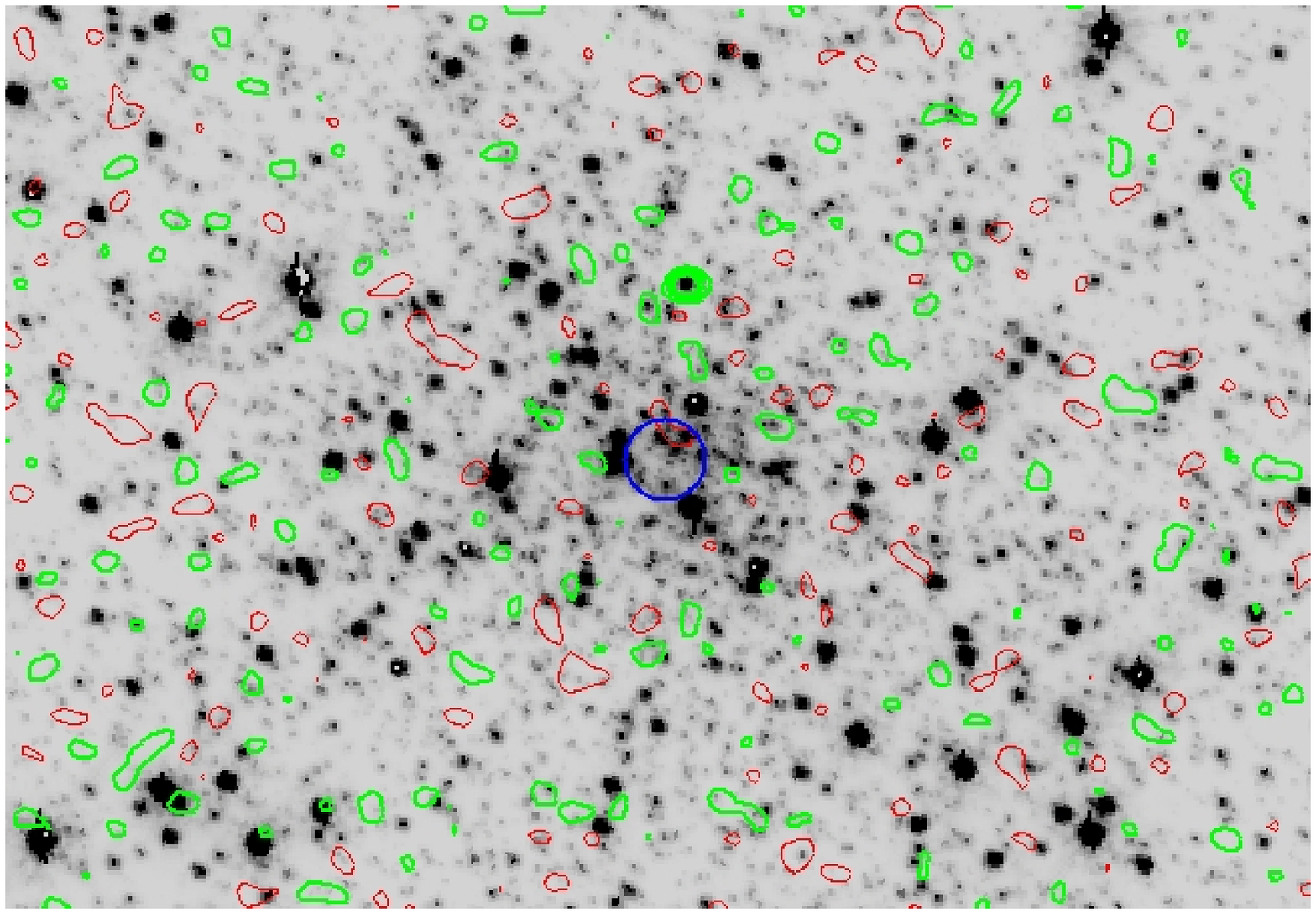,width=9cm,angle=0}
\figcaption[f1.ps]{{\it HST} optical image of M15 overlayed with the
VLA contours of the central 10\arcsec. Positive 1,2, and 3$\sigma$
noise contours (8.5, 17, and 25.5 $\mu$Jy) are shown in green and
negative are shown in red. The blue circle marks the center determined
from Noyola \& Gebhardt (2006), with a diameter of 0.5\arcsec. North
is up and East to the left. The radio source, AC211, is easily seen
just north of the center. The other known X-ray source, M15 X-2, is
not detected.
\label{fig1}}
\vskip 10pt

\vskip 10pt \psfig{file=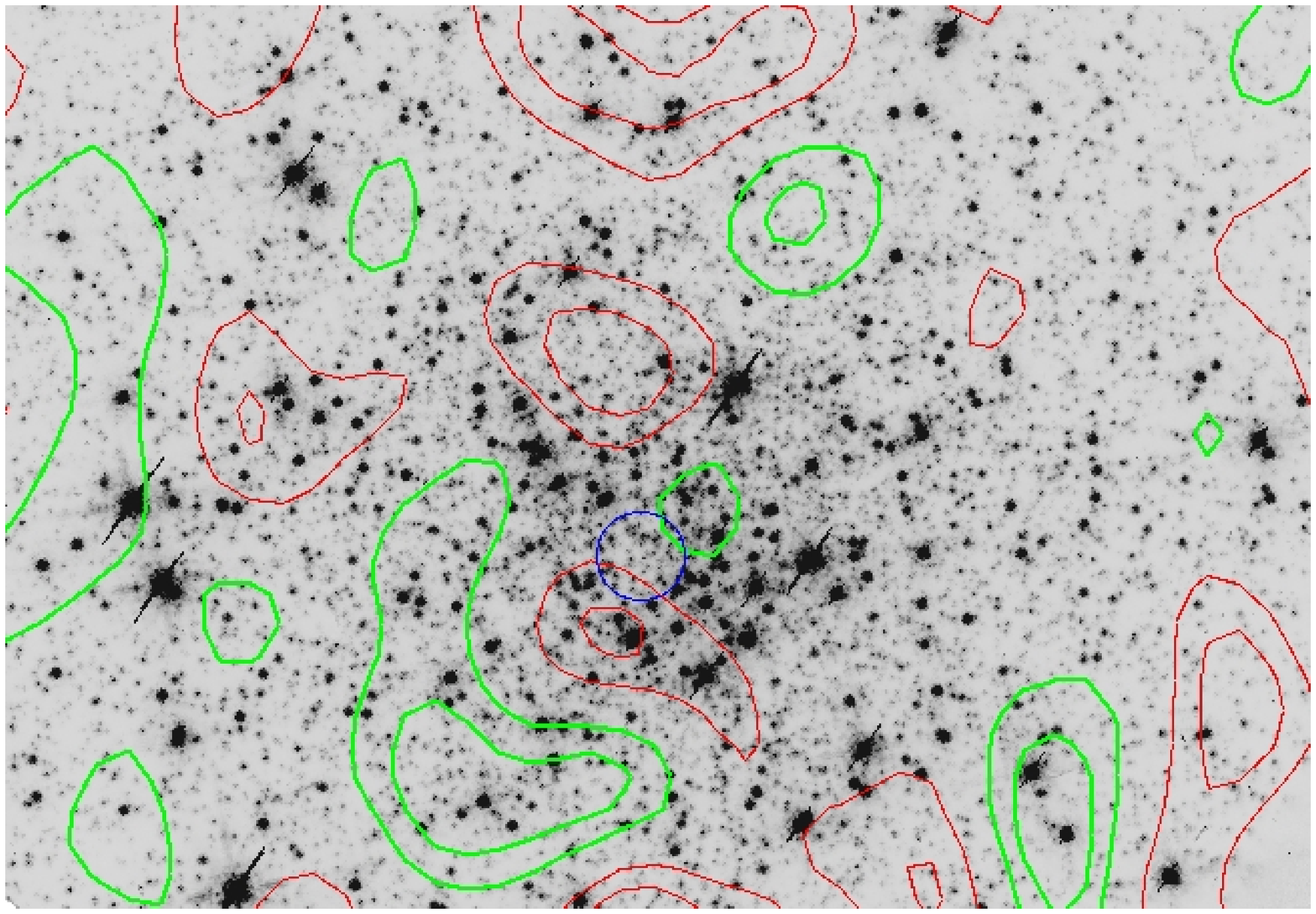,width=9cm,angle=0}
\figcaption[f2.ps]{{\it HST} optical image of NGC~6093 (M80) overlayed
with the VLA contours of the central 10\arcsec. Positive 1,2, and
3$\sigma$ noise contours (12, 24, and 36 $\mu$Jy)are shown in green
and negative are shown in red. The blue circle marks the center
determined from Noyola \& Gebhardt (2006), with a diameter of
1\arcsec. North is up and East to the left.
\label{fig2}}
\vskip 10pt

\vskip 10pt \psfig{file=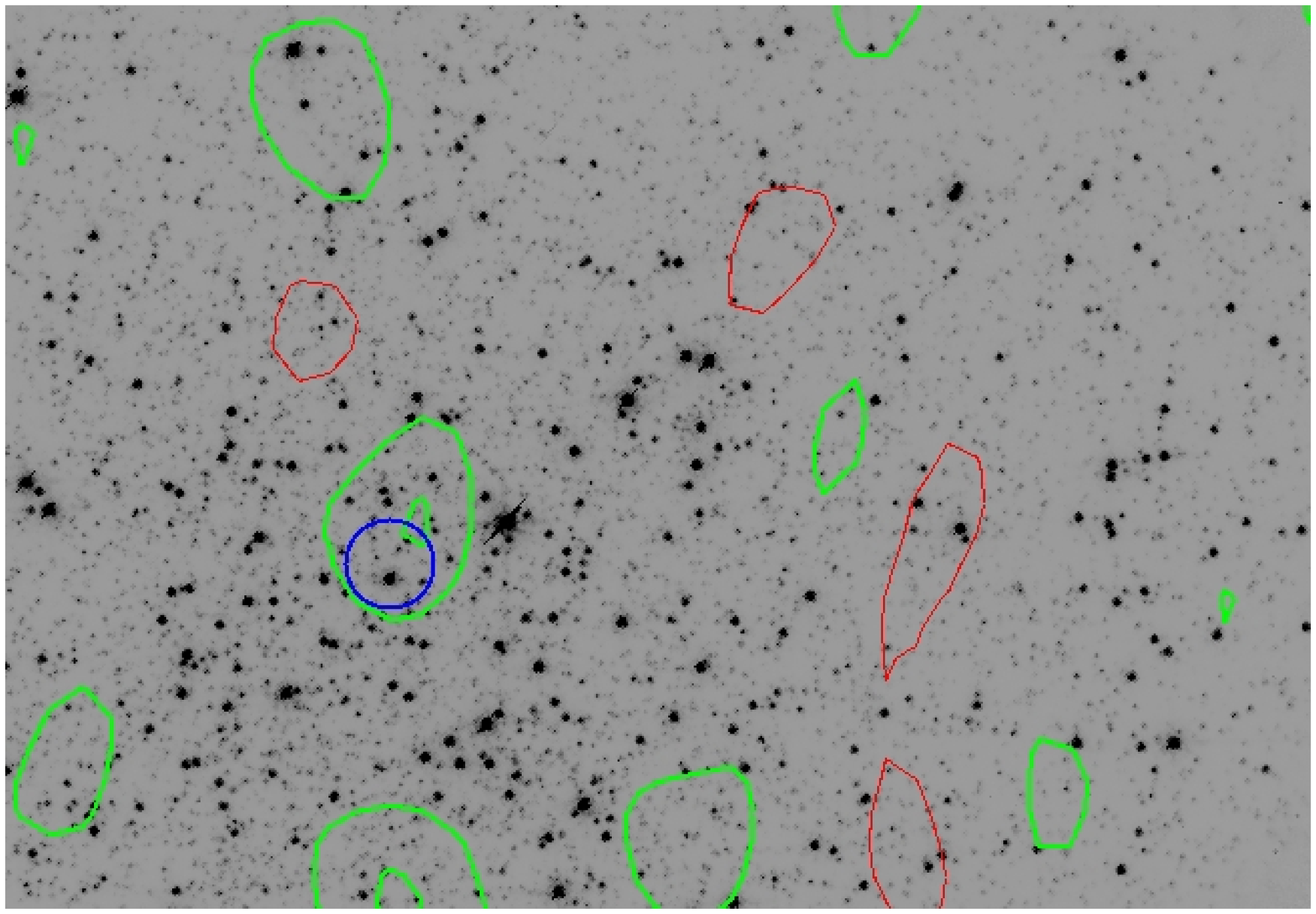,width=9cm,angle=0}
\figcaption[f3.ps]{{\it HST} optical image of NGC~6266 (M62)
overlayed with the VLA contours of the central 10\arcsec. Positive
1,2, and 3$\sigma$ noise contours (12, 24, and 36 $\mu$Jy)are shown in
green and negative are shown in red. The blue circle marks the center
determined from Noyola \& Gebhardt (2006), with a diameter of
1\arcsec. North is up and East to the left. There is about a
2-$\sigma$ positive 8Ghz signal at the center.
\label{fig3}}
\vskip 10pt

\section{Discussion}

Failure to detect radio radiation at 8.6 GHz from the centers of three
globular clusters does not prove that no globular clusters have IMBHs
at their centers. Besides not having a black hole, other
interpretations include 1) accretion by the BH could be episodic and
we happened to observe the BHs in an ``off-state", 2) the gas density
could be much lower compared to galaxies, 3) the radiative efficiency
may be lower than assumed (although the assumed efficiencies are
already quite low), 4) or the accretion model may not be adequate in
general. We would predict, using the relation of Merloni et al. (2003)
or using standard accretion models and gas density estimates (as done
in Maccarone 2004), that we should have detected radio radiation at
8.6 GHz if accretion is steady and the accretion rate times the Bondi
rate is 10$^{-4}\times$ or higher.  We would not have been able to
detect the flux density predicted by a rate of 10$^{-5}\times$ or
less. Ulvestad et al. (2007) estimate the fraction of the Bondi rate
of just under 1\% for G1, but it is difficult to interpret due to the
unknown radiative efficiency. For galactic black holes, the radiative
efficiencies appear to vary greatly with some lower than $10^{-5}$
(Lowenstein et al. 2001), although consistent with rates of around
10\% of the Bondi rate.

Models which predict 8.6 GHz flux densities from central BHs in
globular clusters above about 25 $\mu \rm Jy$/beam can be tested with
the VLA currently. The EVLA should produce, for continuum
observations, a sensitivity improvement of about a factor of 15,
making 8.6 GHz flux densities above about 2$\mu\rm Jy$/beam
detectable.

\vfill\eject

\begin{table}
\begin{center}
\caption{8.6 GHz Flux Density Values}
\begin{tabular}{lcccc}
\tableline\tableline
Cluster & M$\rm_{BH}$ & M$\rm_{BH}$ & Distance & Flux Density \\
\tableline
   & Maccarone, M$_{\rm sun}$ & this paper, M$_{\rm sun}$ &
kpc & $\mu$Jy \\
\tableline
  NGC6093 (M80) & \nodata & 1600 &  8 & $2\times10^3-10^5$\\
  NGC6266 (M62) & 450     & 3000 &  6 & $3\times10^3-10^5$\\
  NGC7078 (M15) & 440     & 1000 & 10 & $1\times10^3-10^5$\\
\tableline
\end{tabular}
\end{center}
\end{table}

\begin{table}
\begin{center}
\caption{Observations}
\begin{tabular}{lccccc}
\tableline\tableline
Cluster  & RA & DEC & Integration & Beam; Pos. Ang    & 3$\sigma$ Limit\\
 & J2000 & J2000 & hours & arcseconds; deg & $\mu$Jy \\
\tableline
  NGC6093 &  16:17:05.00 & $-$22:59:47.3 & 7.0 & $3.9\times2.3$;  -6 & 36\\
  NGC6266 &  17:01:12.96 & $-$30:06:46.2 & 7.0 & $4.7\times2.2$;  -6 & 36\\
  NGC7078 &  21:29:58.35 & $+$12:10:01.5 & 6.5 & $0.2\times0.2$; -76 & 25 \\
\tableline
\end{tabular}
\end{center}
\end{table}


\begin{references}

\reference{} Barth, A., Greene, J., \& Ho, L.C. 2005, \apjl, 619, L151
\reference{} Baumgardt, H., Hut, P., Makino, J., McMillan, S., \&
Portegies Zwart, S. 2003a, \apjl, 582, L21
\reference{} Baumgardt, H., Makino, J., Hut, P., \& Portegies Zwart,
S. 2003b, \apjl, 589, L25
\reference{} Colpi, M., Mapelli, M., \& Possenti, A. 2003, \apj, 599, 1260
\reference{} De Rijcke, S., Buyle, P., \& Dejonghe, H. 2006, \mnras, 368, 43
\reference{} Ebisuzaki, T., et al. 2001, ApJ, 562, L19
\reference{} Ferrarese, L., \& Merritt, D. 2000, \apjl, 539, L9
\reference{} Freire, P., Kramer, M., Lyne, A., Camilo, F., Manchester, R.
\& D'Amico, N. 2001, \apjl, 557, L105
\reference{} Gebhardt, K., et al. 2000, \apj, 539, L13
\reference{} Gebhardt, K., Rich, R.M.R., \& Ho, L.C. 2002, \apjl, 578, L41
\reference{} Gebhardt, K., Rich, R.M.R., \& Ho, L.C. 2005, \apj, 634, 1093
\reference{} Greene, J. \& Ho, L.C. 2006, \apjl, 641, L21
\reference{} Gurkan, M., Freitag, M. \& Rasio, F. 2004, \apj, 604, 632
\reference{} Haehnelt, M.~G., \& Kauffmann, G. 2000, \mnras, 318, L35
\reference{} Illingworth, G. \& King, I. 1977, \apjl, 218, L109
\reference{} Jonhston, H., Kulkarni, S., \& Goss, W.M. 1991, \apjl, 382, L89
\reference{} Lowenstein, M., Mushotsky, R., Angelini, L., Arnaud, K.,
\& Quataert, E. 2001, \apjl, 555, L21
\reference{} Maccarone, T. 2004, MNRAS, 351, 1049
\reference{} Maccarone, T. \& Koerding 2006, Astronomy \& Geophysics, 47, 29
\reference{} Maccarone, T., Fender, R., \& Tzioumis, A. 2005, Ap\&SS, 300, 247
\reference{} McLaughlin, D., Anderson, J., Meylan, G., Gebhardt, K.,
Pryor, C., Minniti, D., \& Phinney, S. 2006, \apjs, 166, 249
\reference{} Merloni, A., Heinz, S., \& di Matteo, T. 2003, \mnras, 345, 1057
\reference{} Miller, M.C., \& Hamilton, D.P. 2002, MNRAS, 330, 232
\reference{} Newell, B., Da Costa, G., \& Norris, J. 1976, \apjl, 208, L55
\reference{} Noyola, E. \& Gebhardt, K. 2006, \aj, 132, 447
\reference{} Pooley, D \& Rappaport, S. 2006, \apjl, 644, L45
\reference{} Portegies Zwart, S., \& McMillan, S. 2002, ApJ, 576, 899
\reference{} Robertson, B., Hernquist, L., Cox, T., Di Matteo, T., Hopkins, P.,
Martini, P., \& Springel, V. 2006, \apj, 641, 90
\reference{} Silk, J., \& Rees, M. J. 1998 A\&A, 331, L1
\reference{} Trudolyubov, S. \& Priedhorsky, W. 2004, \apj, 616, 821
\reference{} Ulvestad, J., Greene, J., \& Ho L.C. 2007, \apjl, 661, 151
\reference{} van der Marel, R.P., Gerssen, J., Guhathakurta, R.,
Peterson, R., \& Gebhardt, K. 2002, AJ, 124, 3255
\reference{} White, N., \& Angelini, L. 2001, \apjl, 561, L101

\end{references}
\end{document}